\documentstyle[11pt]{article}
   \newcommand {\nc}{\newcommand}
   \nc{\eq}{\begin{equation}}
   \nc{\en}{\end{equation}}
   \nc{\eqa}{\begin{eqnarray}}
   \nc{\ena}{\end{eqnarray}}
   \nc{\eqann}{\begin{eqnarray*}}
   \nc{\enann}{\end{eqnarray*}}
   \nc {\dfn}[1]{{\it{#1}}}
   \nc{\nn}{\nonumber}
   \def\dlt{\delta}
   \def\ep{\epsilon}
   
   \def\rep{representation}
   \def\repv{representation }

   \def\etal{{\it et al} }

   \nc{\sqt}{\sqrt{2}}
   \nc{\tsqt}{2\sqt}
   \nc{\msqt}{$\sqt$}
   \nc{\nsqt}{$-\sqt$}
   \nc{\mtsqt}{$\tsqt$}
   \nc{\ntsqt}{$-\tsqt$}

   \def\ot{\otimes}
   
   \def\smdp{>\hspace{-0.2cm}\lhd}

   \nc {\inv}[1]{#1^{-1}}
   \nc {\hc}[1]{{#1}^\dag}
   \nc {\cc}[1]{{#1}^\ast}
   \nc {\ad}[2]{Ad_{#1}({#2})}
   \nc {\wad}{\widetilde{Ad}}
   \nc {\wadf}[2]{\widetilde{Ad}_{#1}({#2})}
   \nc {\pb}[1]{{#1}^\ast}
   \def\tld{\tilde}
   \def\pr{\prime}

   \def\intg{{\cal Z}}
   
   \def\complex{{\cal C}}
   
   \def\Alg{{\cal A}}
   \def\Hs{{\cal H}}

   \def\Drc{{\cal D}}
   \nc {\ga}[2]{{#1}[{#2}]}
   \nc {\cga}[1]{\ga{\complex}{#1}}
   \nc {\cgag}{\cga{G}}
   \nc {\eu}[1]{E^{#1}}
   \nc {\euf}{\eu{4}}
   \nc {\eun}{\eu{n}}
   \nc {\zn}[1]{\intg^{#1}}
   \nc {\zf}{\zn{4}}
   \nc {\zt} {Z_2}
   \nc {\ztn}[1]{\zt^{#1}}
   \nc {\ztt} {\ztn{2}}
   \nc {\ztth} {\ztn{3}}
   \nc {\ztf} {\ztn{4}}
   \nc {\ztnf}{\ztn{n}}
   \nc {\per}[1]{S_{#1}}
   \nc {\pern}{\per{n}}
   \nc {\irrr}[2]{IRR_{#1}({#2})}
   \nc {\irrc}[1]{\irrr{\complex}{#1}}
   \nc {\irrcg}{\irrc{G}}

   \nc {\mn}{(-)}
   \nc {\mns}[1]{\mn^{#1}}
   \nc {\mo}{(-1)}
   \nc {\mos}[1]{\mo^{#1}}
   \nc {\unit} {{\bf 1}}
   \nc {\unt} {\unit_{2\times 2}}
   \nc {\unth} {\unit_{3\times 3}}
   \nc {\unf} {\unit_{4\times 4}}
   \nc {\une} {\unit_{8\times 8}}
   \nc {\unw} {\unit_{12\times 12}}
   \nc {\zrt} {{\bf 0}_{2\times 2}}
   \nc {\zrth} {{\bf 0}_{3\times 3}}
   \nc {\zrf} {{\bf 0}_{4\times 4}}
    \def\CDalign#1{\bgroup\vcenter\bgroup\tabskip 2pt 
      \baselineskip 14pt \lineskip 3pt \lineskiplimit 3pt
      \halign\bgroup &\hfill$##$\hfill\crcr
      #1\crcr\egroup\egroup\egroup} 
    
  \nc{\act}[3]{S_{{#1}}^{{#2}}[{#3}]}
  \nc{\cact}[2]{S_{Cl}^{{#1}}[{#2}]}
  
   \nc {\prj}[4]{\pi_{#1,#2}#2\ot_{S_#1}e^o_{#3,#4}}
   \nc {\prjf}[2]{\prj{o}{#1}{\eta}{#2}}
   \nc {\prjff}{\prjf{h}{i}}
   \nc {\Prj}[4]{\Pi_{#1,#2;#3,#4}}
   \nc {\Prjf}[2]{\Prj{o}{#1}{\eta}{#2}}
   \nc {\Prjff}{\Prjf{h}{i}}
   \nc {\orbt}[2]{\pi_{#1,#2}}
   \nc {\orbtf}[1]{\orbt{o}{#1}}
   \nc {\orbte}{\orbtf{e}}
   \nc {\rp}[3]{\orbt{#1}{#2}(#3)}
   \nc {\rpf}[2]{\rp{o}{#1}{#2}}
   \nc {\stbrp}[2]{D^o_#1(\tld{s}(#2))}
   \nc {\stbrpme}[4]{\stbrp{#1}{#2}^{#3}_{#4}}
   \nc {\stbrpf}[1]{\stbrp{\eta}{#1}}
   \nc {\stbrpmef}[3]{\stbrpme{\eta}{#1}{#2}{#3}}
   \nc {\ch}[2]{\chi_{#1;#2}}
   \nc {\che}[3]{\ch{#1}{#2}(#3)}
   \nc {\chf}[1]{\ch{o}{#1}}
   \nc {\chff}{\chf{\eta}}
   \nc {\chef}[2]{\che{o}{#1}{#2}}
   \nc {\cheff}[1]{\chef{\eta}{#1}}
   \nc {\chsb}[1]{\chi^o_{#1}}
   \nc {\chsbe}[2]{\chsb{#1}(#2)}
   \nc {\chsbf}{\chsb{\eta}}
   \nc {\chsbef}[1]{\chsbe{\eta}{#1}}
   \nc {\cg}[1]{O_{#1}}
   \nc {\cgn} {\cg{n}}
   \nc {\oh} {\cg{4}}
   \nc {\ohd} {{\overline{\oh}}}
   \nc {\ztfb} {\overline{\ztf}}
   \nc {\iso}[2]{ISO_{#1}(#2)}
   \nc {\isod}[1]{\iso{d}{#1}}
   \nc {\eg}[1]{ISO(#1)}
   \nc {\egn}{\eg{n}}
   \nc {\fs}[1]{F(#1)}
   \nc {\fsf}{\fs{\emb}}
   \nc {\fx}[1]{I(#1)}
   \nc {\fxf}{\fx{\emb}}
   \nc {\wrn}[1]{\ztn{#1}\smdp\per{#1}}
   \nc {\wn}{\wrn{n}}
   \nc {\wnn}{\pern^{\zt}}
   \nc {\w}[1]{\per{#1}^{\zt}}
   \nc {\fix}[1]{\per{(n-#1)}\ot \per{p}}
   \nc {\fixp}{\fix{p}}
   \nc {\cb}[1]{C_{#1}}
   \nc {\cn}{\cb{n}}

   \nc {\prm}{\sigma}
   \nc {\prmt}{\tld{\prm}}
   \nc {\prmp}{\prm^\pr}
   \nc {\cyc}[2]{\tau_{{#1}{#2}}}
   \nc {\emb}{\iota}
   \nc {\epy}{\tld{\ep}}
   \nc {\de}[1]{d_{E^{#1}}}
   \nc {\den}{\de{n}}
   \nc {\dy}{\tld{d}}
   \nc {\repw}[2]{e_{(#1)#2}}
   \nc {\repww}[4]{\repw{#1}{#2}\ot\repw{#3}{#4}}
   \nc {\prw}[6]{\pi_{#1,#2}#2\ot_{F_#1}(\repww{#3}{#4}{#5}{#6})}
   \nc {\dm}[1]{d_{({#1})}}
   \nc {\sls}{{\it slash}}
   \nc {\slsv}{{\it slash} }
   \nc {\lcl} {{\it local}}
   \nc {\lclv} {{\it local} }
   
   \nc{\op}{orientation-preserved}
   \nc{\opv}{orientation-preserved }
   \nc{\on}[1]{SO_{#1}}
   \nc{\ohn}[1]{O_{#1}}
   \nc{\of}{\on{4}}
   \nc{\ohf}{\ohn{4}}
   \nc{\ofd}{\overline{\of}}
   \nc{\ohfd}{\overline{\ohf}}
   \nc{\cir}[1]{e^{i({#1})\pi}}

   \def\dop{Dirac operator}
   \def\dopv{Dirac operator }

   \nc {\norm}[1]{\parallel{#1}\parallel}
   
   \def\DCv{Dirac-Connes }
   \nc{\groupsum}[1]{{\sum_{#1}}^{\pr}}
   \nc{\reducedsum}[1]{{\sum_{#1}}^{\pr\pr}}
 \title{Wilson Action of Lattice Gauge Fields with An Additional Term from Noncommutative Geometry
 }
 \author{
  Jian Dai$^\dag$, Xing-Chang Song\\
  Theoretical Group, Department of Physics, Peking University, Beijing, P.R.China, 100871\\
  \dag Room 2082, Building 48, Peking University, Beijing, P. R. China,
  100871\\
  E-mail: daijianium@yeah.net, songxc@ibm320h.phy.pku.edu.cn
  }
 \date{January 23th, 2001}
 
\begin{document}
  \begin{titlepage}
  \maketitle
  \noindent
  \begin{abstract}
   Differential structure of lattices can be defined if the lattices are treated as models of noncommutative
   geometry. The detailed construction consists of specifying a generalized
   \dopv and a wedge product. Gauge potential and field strength tensor can be
   defined based on this differential structure. When an inner
   product is specified for differential forms, classical action
   can be deduced for lattice gauge fields. Besides the familiar
   Wilson action being recovered, an additional term, related to
   the non-unitarity of link variables and loops spanning no area, emerges.
   \\


  \end{abstract}
  \end{titlepage}
  \section{Introduction}
   Our work gains inspiration from three sources. First, exterior algebra, the foundation of differential forms on differentiable
   manifolds, has an intimate relation with Clifford algebra, the root of spinor \cite{harvey}. Moreover, exterior
   differential together with metric on differential manifolds can be realized strikingly
   on the spinor bundles by \dopv under A. Connes' operator theoretic construction of classical geometry \cite{rennie}.
   Physically speaking the same results, fermions provide
   \rep s for differential structures as well as metric structures of manifolds \cite{Connes1}\cite{schucker}. Connes generalized
   this scenario to noncommutative algebra settings and introduced noncommutative geometry(NCG) \cite{Connes2}.
   Second, a lattice provides a marvelous model for NCG
   developed by Dimakis and M\"{u}ller-Hoissen(D\&M) \cite{dimakis}. D\&M geometry on discrete sets is essentially
   cohomologic description of broken lines; there is no non-commutativity at the level of 0-step broken lines
   which are just functions on these sets, while the non-commutativity is characterized by that the ordering to combine two broken lines not all of 0-step
   can not be exchanged.
   Third, a proper lattice \dop, with or without root in NCG, more or less being attempted to
   solve the problem of chiral fermion in lattice field theory(LFT), was pursued by different authors
   \cite{fls}\cite{viz}\cite{Japan}.\\

   We devised a \dopv $\Drc$ on a n-dimensional lattice recently with the following properties:
   i) Adopting {\it Connes' distance formula}, we prove that the {\it induced
   metric of $\Drc$} recovers Euclidean geometry on this lattice,
   when $n=1,2$ \cite{ds1}; ii) $\Drc$ is a {\it Fredholm operator of index zero}\cite{mad}, i.e. an operator subjected to a
   ``{\it geometric square-root}'' condition $\Drc^2\sim \unit$ \cite{ds2}; iii) A differential structure on the lattice defined
   by {\it reduction of calculus} within the formalism of D\&M can be realized onto spinor space upon this lattice
   through $\Drc$, what's more this \repv is {\it Junk-free} \cite{ds3}(see \cite{va} for a mathematical reference);
   iv) Under a specific matrix \rep, $\Drc$ possesses a
   {\it staggered \dopv}interpretation, if $\Drc$ is slightly modified to fit a ``{\it physical square-root}'' condition
   $\Drc^2=\Delta$ where $\Delta$ is the lattice laplacian \cite{ds3}.
   Till now we just explored properties of this operator on a lattice without presenting of gauge fields.
   In this contribution, gauge coupling is added within the conventional approach in Connes' geometry. We will show
   that the unitarity of \repv of gauge groups results from the compatibility of involution of covariant differential
   and gauge transformations. The classical action of lattice gauge fields is calculated. Resulting from the differential structure defined by us,
   an additional term, which will vanish under the usual assumption of unitarity of link variables in LFT, appears aside to
   the familiar Wilson action for lattice gauge fields.
   The losing of unitarity of link variables will be showed having effect on the induced metric on lattices.
   Other authors' works with similar purpose deserve to be referred here \cite{DMa1}\cite{lsw}, though due to the difference of differential structure
   definitions they did not notice the additional term.\\

   This paper is organized as following.
   In Section \ref{sec1}, necessary notations and formalism of differential forms on a lattice are introduced
   for describing lattice gauge fields,
   and then field strength tensor is calculated.
   In Section \ref{sec2},
   a inner product on differential forms is defined and the classical action of lattice gauge fields,
   which differs from the conventional Wilson action by a term vanishing if link variables are unitary, is calculated.
   Finally we discuss the significance of non-unitary link variables and
   the mathematical rigidity of our work in Section \ref{sec3}.
  \section{Differential Forms, Connection and Curvature on Lattices}
  \label{sec1}
   {\bf Notations}
   A n-dimensional lattice can be regarded as a discrete abelian group
   $\intg^n_N$, the direct product of n cyclic groups $\intg_N$,
   in which $N$ is an integer being large enough. The elements in
   $\intg_N$ can be written as $0,1,...,N-1$ and the
   multiplication is just the addition modulo $N$.
   Let $\Alg$ be the algebra of complex functions on $\intg^n_N$.
   A collection of delta functions on $\intg^n_N$ is defined to be a subset of
   $\Alg$: $\epsilon^x(y)=\Pi_{i=1}^n\delta^{x^i}_{y^i}, \forall x,y \in\intg^n_N$
   where $x^i,y^i$ are the ith components of $x,y$ respectively.
   The algebraic structure of $\Alg$ can be expressed as
   $\epsilon^x\epsilon^y=\Pi_{i=1}^n\dlt^{x^iy^i}$, $\sum_x\epsilon^x=\unit_\Alg$ where $\unit_\Alg(x)=1$,
   since all $\epsilon^x$ form a natural basis for $\Alg$.
   Shift operators acting on $\Alg$ is defined by $(T^\pm_\mu
   f)(x)=f(x\pm \hat{\mu}), \forall x\in \intg^n_N, \forall f\in\Alg$ or equivalently $T^\pm_\mu \epsilon^x=\epsilon^{x\mp\hat{\mu}}$,
   where $\hat{\mu}$ denotes the unit vector along
   $\mu$-axis. Introduce formal partial derivatives by
   $\partial^\pm_\mu:=T^\pm_\mu-\unit$ with $\unit$ being the trivial action on
   $\Alg$. $End_\complex(V)$ refers to $\complex$-linear endomorphism
   algebra on a linear space $V$, thus $T^\pm_\mu, \unit, \partial^\pm_\mu$ are
   all elements in $End_\complex(\Alg)$.
   \subsection{Differential Structure on Lattices}
   Based on the observation that $(\partial^\pm_\mu f)(x)$ are two completely independent numbers with $x$
   being fixed on a lattice, we introduce differential of a function $f$ in $\Alg$ in this way
   \[
    df=\sum_{\mu=1}^n(\partial^+_\mu f\chi^\mu_++\partial^-_\mu f\chi^\mu_-)
   \]
   in which $\chi^\mu_\pm$ form a basis of {\it first-order
   differential forms}. This intuition gains its rigidity under the following construction.
   $\Alg$ can be extended to be a graded differential algebra
   $(\Omega(\Alg),d)$ by the construction rules shown below:
   $i)\Omega(\Alg)=\oplus^\infty_{k=0}{\Omega^k(\Alg)}$, $\Omega^0(\Alg)=\Alg$ and the elements in $\Omega^p(\Alg)$ are
   referred as p-order differential forms or just p-forms;
   $ii)\Omega^p(\Alg)\cdot\Omega^q(\Alg)=\Omega^{p+q}(\Alg)$;
   $iii)d:\Omega^k(\Alg)\rightarrow\Omega^{k+1}(\Alg), k=0,1,...$
   is a linear map satisfying graded Leibnitz rule and nilpotent rule
   \[
    d(\omega_p\omega^\pr)=d(\omega_p)\omega^\pr+(-)^p\omega_pd(\omega^\pr),
    \forall \omega_p\in\Omega^p(\Alg),\omega^\pr\in\Omega(\Alg),
   \]
   \[
    d^2=0;
   \]
   iv)$\unit_\Alg$ is the unit of $\Omega(\Alg)$. To make the construction be consistent,
   some conditions on $\chi^\mu_\pm$ have to be imposed,
   \eq\label{n1}
    \chi^\mu_\pm f=(T^\pm_\mu f)\chi^\mu_\pm
   \en
   \eq\label{c1}
    \{\chi^\mu, \chi^\nu\}=0, \{\chi^{-\mu}, \chi^{-\nu}\}=0,
    \{\chi^\mu, \chi^{-\nu}\}=\dlt^{\mu\nu}d\chi^\mu=\dlt^{\mu\nu}d\chi^{-\nu}
   \en
   for all $\mu,\nu=1,2,...,n$. We refer \cite{sitarz}\cite{ds3}
   for a detailed account of these statements. Eq.(\ref{n1}) is the fundamental non-commutativity on lattices,
   which indicates that functions is no longer commutative with differential
   forms, while the ``error'' of this non-commutativity is a shift by one lattice spacing, hence vanishing
   under {\it continuum limit}.
   The geometric interpretation of above construction is
   that p-forms correspond to linear combinations of p-stepped broken lines on $\intg^n_N$. For example, let $n=1$ and consider
   $\chi_+$. It is easy to show that $\chi_+=\sum_{x=0}^{N-1} \epsilon^xd\epsilon^{x+1}$, thus $\chi_+$ can be interpreted as combination of
   1-step line segments from $x$ to $x$+1 with coefficients all being one.\\

   Next we formulate a noncommutative exterior differential algebra $(\Lambda(\Alg), d)$ by
   introducing an equivalent relation $d\chi^\mu_\pm\sim 0$ onto
   $(\Omega(\Alg),d)$, s.t. $\Lambda(\Alg)\cong\Omega(\Alg)/\sim$.
   This definition avoids the potential ambiguity of applying wedge product directly, which results from
   non-commutative relation Eq.(\ref{n1}).
   Equivalently and practically, $\Lambda(\Alg)$ can be defined to be a subset of $\Omega(\Alg)$
   together with a projection $\Pi$ such that
   $\Lambda(\Alg)=\Pi(\Omega(\Alg))$, in which $\Pi$ is defined as
   \eq\label{Pi}
    \Pi(f_0\chi^\mu_sf_1\chi^\nu_tf_3):=
    {1\over 2}f_0(T^s_\mu f_1)(T^s_\mu T^t_\nu
    f_3)(\chi^\mu_s\chi^\nu_t-\chi^\nu_t\chi^\mu_s)=f_0(T^s_\mu f_1)(T^s_\mu T^t_\nu f_3)\chi^\mu_s\wedge\chi^\nu_t
   \en
   for all $f_\alpha\in\Alg,\alpha=0,1,2$, $s,t\in\{+,-\}$,
   $\mu,\nu=1,2,...,n$. Note that $\Pi(f_0\chi^\mu_s\cdot f_1\chi^\nu_t)=f_0(T^s_\mu
   f_1)\chi^\mu_s\wedge\chi^\nu_t\neq
   -(T^t_\nu
   f_0)f_1\chi^\mu_s\wedge\chi^\nu_t=\Pi(f_1\chi^\nu_t\cdot
   f_0\chi^\mu_s)$ generally, therefore $\omega_{(1)}\wedge\omega_{(1)}$ is
   not necessarily equal to zero for a generic 1-form $\omega_{(1)}$.
   The notion of wedge product makes sense by the projection $\Pi$, and
   $(\Lambda(\Alg),d)$ will be referred as a differential structure
   on $\intg^n_N$.
   \subsection{Representation of Differential Structure,
    Dirac-Connes Operator on the Lattice}
   A ``fermion'' \repv for $(\Lambda(\Alg),d)$ is developed in this subsection.
   Introduce a spinor space $\Hs_s=\complex^{2^n}$, and let $\Hs$ be a finite dimensional Hilbert space $\Alg\ot
   \Hs_s$ under the conventional definition of inner
   product
   $(\psi_1,\psi_2)=\sum_{x\in\intg^n}\sum_{i=1}^{2^n}\overline{\psi_1^i(x)}\psi_2^i(x)$. $\Alg$
   is represented on $\Hs$ by $\pi:\pi(f)=f\ot \unit_s$, thus $\Hs$ is turned out to be a left free $\Alg$-module
   in mathematical literature.
   Now extend $\pi$ to be the ``fermion'' \repv on $End_\complex(\Hs)$ by specifying $\pi(\chi^\mu_\pm)$ and applying algebraic homomorphism rule.
   First, define
   \eq\label{def}
    \pi(\chi^\mu_\pm)=i\eta^\mu_\pm,
    \eta^\mu_\pm=T^\pm_\mu\ot\Gamma^\mu_\pm
   \en
   $\Gamma^\mu_\pm$ are generators of Clifford algebra of
   2n-dimensional Euclidean space, thus satisfy that
   \[
    \{\Gamma^\mu_+,\Gamma^\nu_+\}=0,
    \{\Gamma^\mu_-,\Gamma^\nu_-\}=0,
    \{\Gamma^\mu_+,\Gamma^\nu_-\}=\dlt^{\mu\nu}\unit_s, \mu,\nu=1,2,...,n
   \]
   as well as that
   $(\Gamma^\mu_\pm)^\dag=\Gamma^\mu_\mp$.
   Accordingly,
   \[
    \{\eta^\mu_+,\eta^\nu_+\}=0,
    \{\eta^\mu_-,\eta^\nu_-\}=0,
    \{\eta^\mu_+,\eta^\nu_-\}=\dlt^{\mu\nu}\unit\ot\unit_s,
    \mu,\nu=1,2,...,n
   \]
   which is the \repv of Eq.(\ref{c1}). Since
   $(T^\pm_\mu)^\dag=T^\mp_\mu$, there is
   $(\eta^\mu_\pm)^\dag=\eta^\mu_\mp$.
   Eq.(\ref{n1}) is realized as $\eta^\mu_\pm \pi(f)=\pi(T^\pm_\mu
   f)\eta^\mu_\pm$ and $\pi(df)=i[\Drc,\pi(f)]$ where
   \[
    \Drc:=\sum_{\mu=1}^n(\eta^\mu_++\eta^\mu_-)
   \]
   $\Drc$ is a so-called \DCv operator in NCG. One can check that
   the relation $\Drc^2=n\unit\ot\unit_s$ holds, which is the condition for a {\it Fredholm operator of index
   zero}. Consequently, $(\Hs,\Drc)$ forms a {\it Fredholm module}
   on $\Alg$. Besides, it is obvious
   that $\Drc$ is hermitian, $\Drc^\dag=\Drc$. Second,
   Eq(\ref{Pi}) is implemented by defining the product of two
   conjoint $\eta^\mu_s,\eta^\nu_t$ to be wedge product, thus a formal noncommutative $\Alg$-linear rule
   $\eta^\mu_s\wedge(\eta^\nu_t f)=\eta^\mu_s\wedge ((T_\nu^t
   f)\eta^\nu_t)
   =(\eta^\mu_s(T_\nu^t f))\wedge\eta^\nu_t=((T_\mu^s T_\nu^t f)\eta^\mu_s)\wedge\eta^\nu_t$ follows.
   \subsection{Gauge Fields on Lattices}
   Let $\complex^k$ be the color space $\Hs_c$ upon which gauge group $G$ is represented as $R:G\rightarrow Aut_\complex (\Hs_c)$.
   Directly product $\Hs$ with $\Hs_c$ to form a new
   $\Alg$-module $\tld{\Hs}=\Hs_c\ot \Hs$. $\pi(\Lambda(\Alg))$
   acts on $\tld{\Hs}$ as $\unit_c\ot \omega, \forall\omega\in \pi(\Lambda(\Alg))$ to which we will
   still use the symbol $\omega$. Let $\tld{\Lambda}(\Alg)=End_\complex(\Hs_c)\ot \pi(\Lambda(\Alg))$
   consisting of {\it differential forms valued in $End_\complex(\Hs_c)$} which are still referred as differential forms without introducing
   confusion.
   A sequence of concepts in gauge theory can be developed following
   a conventional routine.
   Define {\it connection 1-form} to be
   \[
    A=i\sum_{\mu=1}^n(A^+_\mu\eta^\mu_+ + A^-_\mu\eta^\mu_-) \in \tld{\Lambda}(\Alg)
   \]
   in which gauge potential $A^\pm_\mu$ act trivially on spinor
   space $\Hs_s$. $A$ is required to be anti-hermitian,
   $A^\dag=-A$, hence satisfying that
   $A^-_\mu=T_\mu^-(A^{+\dag}_\mu)$. Let {\it covariant
   differential} be $\nabla:=i\Drc+A$, so $\nabla^\dag=-\nabla$.
   Introduce {\it parallel transport} by $U^\pm_\mu=\unit+
   A^\pm_\mu$, then
   \[
    \nabla=i\sum_{\mu=1}^n(U^+_\mu\eta^\mu_+ + U^-_\mu\eta^\mu_-)
   \]
   and
   \eq\label{her}
    U^-_\mu=T_\mu^-(U^{+\dag}_\mu)
   \en
   Note importantly that $U^\pm_\mu$ is referred as link variables in physical literature and that, however, no unitarity
   as a prescription is forced on $U^\pm_\mu$ in this article.
   {\it Curvature} or {\it field strength tensor} is defined
   to be $F=-\nabla\wedge\nabla=-i\{\Drc, A\}_\wedge-A\wedge A$.
   It is easy to check that $F^\dag=F$ and that {\it Bianchi identity} holds for $F$, namely $i[\nabla,
   F]_\wedge=0$. Even if the gauge group is abelian,
   there is still an $A^2$ term in curvature due to
   non-commutativity. The detailed form of $F$ can be computed using either gauge potentials $A^\pm_\mu$
   or parallel transports $U^\pm_\mu$. After a sheet of paper's
   algebra, one will reach that
   $F=\sum_{\mu\nu}(F^{++}_{\mu\nu}\eta^\mu_+\wedge\eta^\nu_++F^{+-}_{\mu\nu}\eta^\mu_+\wedge\eta^\nu_-+
   F^{-+}_{\mu\nu}\eta^\mu_-\wedge\eta^\nu_++F^{--}_{\mu\nu}\eta^\mu_-\wedge\eta^\nu_-)$
   in which
   \eq\label{f1}
    F^{++}_{\mu\nu}={1\over 2}((\partial^+_\mu A^+_\nu -\partial^+_\nu
    A^+_\mu)+(A^+_\mu(T^+_\mu A^+_\nu)-A^+_\nu(T^+_\nu A^+_\mu)))={1\over 2}(U^+_\mu(T^+_\mu U^+_\nu)-U^+_\nu(T^+_\nu
    U^+_\mu))
   \en
   \eq\label{f2}
    F^{+-}_{\mu\nu}={1\over 2}((\partial^-_\mu A^+_\nu -\partial^-_\nu
    A^+_\mu)+(A^+_\mu(T^+_\mu A^-_\nu)-A^-_\nu(T^-_\nu A^+_\mu)))={1\over 2}(U^+_\mu(T^+_\mu U^-_\nu)-U^-_\nu(T^-_\nu
    U^+_\mu))
   \en
   and $F^{--}_{\mu\nu}$, $F^{-+}_{\mu\nu}$ can be given by
   +$\leftrightarrow$- in $F^{++}_{\mu\nu}$,
   $F^{+-}_{\mu\nu}$. We point out that the exchange of $\pm$ is a
   symmetry in our formalism which we will use broadly below, and that $F^{++}_{\mu\nu}=-F^{++}_{\nu\mu}$,
   $F^{--}_{\mu\nu}=-F^{--}_{\nu\mu}$,
   $F^{-+}_{\mu\nu}=-F^{+-}_{\nu\mu}$.\\

   Gauge transformations are 0-forms valued in $R(G)$, denoted by
   $g$. If connect 1-form transforms affinely as
   $A^\pr=gAg^{-1}+ig[\Drc,g^{-1}]$, then covariant differential
   and curvature transform adjointly as $\nabla^\pr=g\nabla
   g^{-1}, F^\pr=gFg^{-1}$. Involution of covariant differential and curvature is
   compatible with any gauge transformation $g$, i.e.
   $\nabla^{\pr\dag}=-\nabla^\pr,F^{\pr\dag}=F^\pr$, if $[\nabla,
   g^\dag g]=0$. Consequently, $g^\dag g$ equals to unit up to an
   overall scalar factor; $R$ becomes a unitary \repv if this factor equals to
   one. Nevertheless, in our understanding, the unitarity of \repv $R$
   does not imply necessarily the unitarity of link variables $U^\pm_\mu$.
  \section{Inner Product on Differential Forms and Classical Action of Gauge Fields}
  \label{sec2}
   An inner product of differential forms $(,):\tld{\Lambda}(\Alg)\ot\tld{\Lambda}(\Alg)\rightarrow \complex$
   has to be specified, such that classical action for
   gauge fields on this lattice can be written as
   \eq\label{act}
    S[U]={1\over 2(tr_s\unit_s)}(F,F)={1\over 2(tr_s\unit_s)}(\nabla^2,\nabla^2)
   \en
   First we introduce a {\it hermitian structure}
   $\langle,\rangle :\tld{\Lambda}(\Alg)\ot\tld{\Lambda}(\Alg)\rightarrow
   End_\complex(\Hs_c)\ot End_\complex(\Alg)$, $\tld{\omega}\ot\tld{\omega}^\pr\mapsto tr_s(\tld{\omega}^\dag\tld{\omega}^\pr)$ for all
   $\tld{\omega},\tld{\omega}^\pr\in\tld{\Lambda}(\Alg)$ where $tr_s$ is trace on $End_\complex(\Hs_s)$.
   Then $(\tld{\omega},\tld{\omega}^\pr):=tr_c Sp\langle\tld{\omega},\tld{\omega}^\pr\rangle$ where $Sp$ is the trace on
   $End_{\complex}(\Alg)$ and $tr_c$ is the trace on
   $End_\complex(\Hs_c)$. Gauge invariance is guaranteed naturally
   under this definition of $S[U]$.
   To simplify the calculation of $S[U]$, we rewrite $F$ to be
   \[
    F=\sum_{\mu\nu}(F^{++}_{\mu\nu}\eta^\mu_+\wedge\eta^\nu_++F^{--}_{\mu\nu}\eta^\mu_-\wedge\eta^\nu_-
    +2F^{+-}_{\mu\nu}\eta^\mu_+\wedge\eta^\nu_-)
   \]
   Consider a metric tensor
   \[
    g_{(2)}:=
    \left(
     \begin{array}{ccc}
      \langle\eta^\mu_+\wedge\eta^\nu_+,\eta^\lambda_+\wedge\eta^\rho_+\rangle&
      \langle\eta^\mu_+\wedge\eta^\nu_+,\eta^\lambda_+\wedge\eta^\rho_-\rangle&
      \langle\eta^\mu_+\wedge\eta^\nu_+,\eta^\lambda_-\wedge\eta^\rho_-\rangle\\
      \langle\eta^\mu_+\wedge\eta^\nu_-,\eta^\lambda_+\wedge\eta^\rho_+\rangle&
      \langle\eta^\mu_+\wedge\eta^\nu_-,\eta^\lambda_+\wedge\eta^\rho_-\rangle&
      \langle\eta^\mu_+\wedge\eta^\nu_-,\eta^\lambda_-\wedge\eta^\rho_-\rangle\\
      \langle\eta^\mu_-\wedge\eta^\nu_-,\eta^\lambda_+\wedge\eta^\rho_+\rangle&
      \langle\eta^\mu_-\wedge\eta^\nu_-,\eta^\lambda_+\wedge\eta^\rho_-\rangle&
      \langle\eta^\mu_-\wedge\eta^\nu_-,\eta^\lambda_-\wedge\eta^\rho_-\rangle
     \end{array}
    \right)
   \]
   Due the +$\leftrightarrow$- symmetry,
   $\langle\eta^\mu_-\wedge\eta^\nu_-,\eta^\lambda_+\wedge\eta^\rho_+\rangle=\langle\eta^\mu_+\wedge\eta^\nu_+,\eta^\lambda_-\wedge\eta^\rho_-\rangle$,
   $\langle\eta^\mu_-\wedge\eta^\nu_-,\eta^\lambda_+\wedge\eta^\rho_-\rangle=-\langle\eta^\mu_+\wedge\eta^\nu_+,\eta^\lambda_+\wedge\eta^\rho_-\rangle$,
   $\langle\eta^\mu_-\wedge\eta^\nu_-,\eta^\lambda_-\wedge\eta^\rho_-\rangle=\langle\eta^\mu_+\wedge\eta^\nu_+,\eta^\lambda_+\wedge\eta^\rho_+\rangle$,
   $\langle\eta^\mu_+\wedge\eta^\nu_-,\eta^\lambda_+\wedge\eta^\rho_+\rangle=-\langle\eta^\mu_+\wedge\eta^\nu_-,\eta^\lambda_-\wedge\eta^\rho_-\rangle$.
   Therefore, only
   $\langle\eta^\mu_+\wedge\eta^\nu_+,\eta^\lambda_+\wedge\eta^\rho_+\rangle$,
   $\langle\eta^\mu_+\wedge\eta^\nu_+,\eta^\lambda_+\wedge\eta^\rho_-\rangle$,
   $\langle\eta^\mu_+\wedge\eta^\nu_+,\eta^\lambda_-\wedge\eta^\rho_-\rangle$,
   $\langle\eta^\mu_+\wedge\eta^\nu_-,\eta^\lambda_+\wedge\eta^\rho_-\rangle$,
   $\langle\eta^\mu_+\wedge\eta^\nu_-,\eta^\lambda_-\wedge\eta^\rho_-\rangle$
   need to be computed.
   Some algebra is needed to show
   \[
    \langle\eta^\mu_+\wedge\eta^\nu_+,\eta^\lambda_+\wedge\eta^\rho_+\rangle
    ={1\over 4}(tr_s\unit_s)
    (\dlt^{\mu\lambda}\dlt^{\nu\rho}-\dlt^{\mu\rho}\dlt^{\nu\lambda})\unit_c\ot\unit
   \]
   \[
    \langle\eta^\mu_+\wedge\eta^\nu_-,\eta^\lambda_+\wedge\eta^\rho_-\rangle
    ={1\over 4}(tr_s\unit_s)
    \dlt^{\mu\lambda}\dlt^{\nu\rho}\unit_c\ot\unit
   \]
   \[
    \langle\eta^\mu_+\wedge\eta^\nu_+,\eta^\lambda_+\wedge\eta^\rho_-\rangle=
    \langle\eta^\mu_+\wedge\eta^\nu_+,\eta^\lambda_-\wedge\eta^\rho_-\rangle=
    \langle\eta^\mu_+\wedge\eta^\nu_-,\eta^\lambda_-\wedge\eta^\rho_-\rangle=0
   \]
   Consequently,
   \[
    g_{(2)}={1\over 4}(tr_s\unit_s)
    \left(
     \begin{array}{ccc}
      (\dlt^{\mu\lambda}\dlt^{\nu\rho}-\dlt^{\mu\rho}\dlt^{\nu\lambda})&0&0\\
      0&\dlt^{\mu\lambda}\dlt^{\nu\rho}&0\\
      0&0&(\dlt^{\mu\lambda}\dlt^{\nu\rho}-\dlt^{\mu\rho}\dlt^{\nu\lambda})
     \end{array}
    \right)
    \unit_c\ot\unit
   \]
   Now apply the result of $g_{(2)}$,
   \[
    \langle F,F\rangle =
    \sum_{\mu\nu\mu^\pr\nu^\pr}
    (T^-_\mu T^+_\nu (F_{\mu\nu}^{+-\dag} F_{\mu^\pr\nu^\pr}^{+-})
    \langle\eta^\mu_+\wedge\eta^\nu_-,\eta^{\mu^\pr}_+\wedge\eta^{\nu^\pr}_-\rangle+
   \]
   \[
    T^-_\mu T^-_\nu (F_{\mu\nu}^{++\dag} F_{\mu^\pr\nu^\pr}^{++})
    \langle\eta^\mu_+\wedge\eta^\nu_+,\eta^{\mu^\pr}_+\wedge\eta^{\nu^\pr}_+\rangle+
    T^+_\mu T^+_\nu (F_{\mu\nu}^{--\dag} F_{\mu^\pr\nu^\pr}^{--})
    \langle\eta^\mu_-\wedge\eta^\nu_-,\eta^{\mu^\pr}_-\wedge\eta^{\nu^\pr}_-\rangle)
   \]
   \eq\label{ff}
    ={1\over 2}(tr_s\unit_s)\sum_{\mu\nu}(
    2T^-_\mu T^+_\nu (F_{\mu\nu}^{+-\dag} F_{\mu\nu}^{+-})+
    T^-_\mu T^-_\nu (F_{\mu\nu}^{++\dag} F_{\mu\nu}^{++})+
    T^+_\mu T^+_\nu (F_{\mu\nu}^{--\dag} F_{\mu\nu}^{--}))
   \en
   Substitute detailed expressions in Eqs.(\ref{f1})(\ref{f2}) into
   (\ref{ff}), and notice anti-hermitian condition (\ref{her}),
   \[
    \langle F,F\rangle=W+(tr_s\unit_s)S
   \]
   in which the symbols are defined in the following way:\\
   Wilson term
   \[
    W=-{1\over 4}(tr_s\unit_s)\sum_{\mu\neq\nu}(
    {\cal F^{-+}_{\mu\nu}}+{\cal F^{+-}_{\mu\nu}}
    +{\cal F^{++}_{\mu\nu}}+{\cal F^{--}_{\mu\nu}})
   \]
   \[
    {\cal F^{\mp\pm}_{\mu\nu}}={1\over 2}(P^{\mp\pm}_{\mu\nu}
    +P^{\pm\mp}_{\nu\mu})-\unit_c\ot\unit,
    {\cal F^{\pm\pm}_{\mu\nu}}={1\over 2}(P^{\pm\pm}_{\mu\nu}
    +P^{\pm\pm}_{\nu\mu})-\unit_c\ot\unit, \mu\neq\nu
   \]
   \[
    P^{-+}_{\mu\nu}=U^-_\mu (T^-_\mu U^+_\nu)(T^-_\mu T^+_\nu
    U^+_\mu)(T^+_\nu U^-_\nu),
    P^{+-}_{\mu\nu}=P^{-+}_{\mu\nu}(+\leftrightarrow -), \mu\neq\nu
   \]
   \[
    P^{--}_{\mu\nu}=U^-_\mu (T^-_\mu U^-_\nu)(T^-_\mu T^-_\nu
    U^+_\mu)(T^-_\nu U^+_\nu),
    P^{++}_{\mu\nu}=P^{--}_{\mu\nu}(+\leftrightarrow -), \mu\neq\nu
   \]
   and an additional term
   \[
    S={1\over 8}\sum_{\mu\neq\nu}
    (S^{-+}_{\mu\nu}+S^{-+}_{\nu\mu}+S^{+-}_{\mu\nu}+S^{+-}_{\nu\mu}+S^{++}_{\mu\nu}+S^{++}_{\nu\mu}+S^{--}_{\mu\nu}+S^{--}_{\nu\mu})
    +{1\over 4}\sum_{\mu}(
    S^{++}_\mu+S^{--}_\mu-S^{+-}_\mu-S^{-+}_\mu)
   \]
   \[
    S^{\mp\pm}_{\mu\nu}=\Pi^{\mp\pm}_{\mu\nu}-\unit_c\ot\unit,
    S^{\pm\pm}_{\mu\nu}=\Pi^{\pm\pm}_{\mu\nu}-\unit_c\ot\unit, \mu\neq\nu
   \]
   \[
    \Pi^{-+}_{\mu\nu}=U^+_\nu (T^+_\nu U^-_\mu)(T^-_\mu T^+_\nu U^+_\mu)(T^+_\nu U^-_\nu),
    \Pi^{+-}_{\mu\nu}=\Pi^{-+}_{\mu\nu}(+\leftrightarrow -), \mu\neq\nu
   \]
   \[
    \Pi^{--}_{\mu\nu}=U^-_\nu (T^-_\nu U^-_\mu)(T^-_\mu T^-_\nu U^+_\mu)(T^-_\nu U^+_\nu),
    \Pi^{++}_{\mu\nu}=\Pi^{--}_{\mu\nu}(+\leftrightarrow -), \mu\neq\nu
   \]
   \[
    S^{\mp\pm}_\mu=\Pi^{\mp\pm}_\mu-\unit_c\ot\unit,
    S^{\pm\pm}_\mu=\Pi^{\pm\pm}_\mu-\unit_c\ot\unit
   \]
   \[
    \Pi^{-+}_\mu=U^-_\mu (T^-_\mu U^+_\mu)U^+_\mu(T^+_\mu U^-_\mu), \Pi^{+-}_\mu=\Pi^{-+}_\mu(+\leftrightarrow -)
   \]
   \[
    \Pi^{--}_\mu=U^-_\mu (T^-_\mu U^+_\mu)U^-_\mu(T^-_\mu U^+_\mu), \Pi^{++}_\mu=\Pi^{--}_\mu(+\leftrightarrow -)
   \]
   The geometric interpretations of $P^{st}_{\mu\nu}(x)$, $\Pi^{st}_{\mu\nu}(x)$, $\Pi^{st}_\mu(x)$ are Wilson-loop operators
   of parallel transports illustrated as
   \[
    P^{\mp\pm}_{\mu\nu}(x): x\rightarrow (x\pm\hat{\nu})\rightarrow
   (x\mp\hat{\mu}\pm\hat{\nu})\rightarrow (x\mp\hat{\mu})\rightarrow x
   \]
   \[
    P^{\pm\pm}_{\mu\nu}(x): x\rightarrow (x\pm\hat{\nu})\rightarrow
   (x\pm\hat{\mu}\pm\hat{\nu})\rightarrow (x\pm\hat{\mu})\rightarrow x
   \]
   \[
    \Pi^{\mp\pm}_{\mu\nu}(x): x\rightarrow (x\pm\hat{\nu})\rightarrow
    (x\mp\hat{\mu}\pm\hat{\nu})\rightarrow (x\pm\hat{\nu})\rightarrow x
   \]
   \[
    \Pi^{\pm\pm}_{\mu\nu}(x): x\rightarrow (x\pm\hat{\nu})\rightarrow
    (x\pm\hat{\mu}\pm\hat{\nu})\rightarrow (x\pm\hat{\nu})\rightarrow x
   \]
   where $\mu\neq\nu$, and
   \[
    \Pi^{\mp\pm}_\mu(x): x\rightarrow (x\pm\hat{\mu})\rightarrow
    x\rightarrow (x\mp\hat{\mu})\rightarrow x
   \]
   \[
    \Pi^{\pm\pm}_\mu(x): x\rightarrow (x\pm\hat{\mu})\rightarrow
    x\rightarrow (x\pm\hat{\mu})\rightarrow x
   \]
   The Wilson-loops producing $P^{st}_{\mu\nu}(x)$ span {\it fundamental plaquettes}, while
   those producing $\Pi^{st}_{\mu\nu}(x)$ and
   $\Pi^{st}_\mu(x)$ span no area.
   Involutive properties can be checked as
   \[
    (P^{\mp\pm}_{\mu\nu})^\dag=P^{\pm\mp}_{\nu\mu},
    (P^{\pm\pm}_{\mu\nu})^\dag=P^{\pm\pm}_{\nu\mu}, \mu\neq\nu
   \]
   \[
    (\Pi^{\mp\pm}_{\mu\nu})^\dag=\Pi^{\mp\pm}_{\mu\nu},
    (\Pi^{\pm\pm}_{\mu\nu})^\dag=\Pi^{\pm\pm}_{\mu\nu}, \mu\neq\nu
   \]
   \[
    (\Pi^{\mp\pm}_\mu)^\dag=\Pi^{\pm\mp}_\mu,
    (\Pi^{\pm\pm}_\mu)^\dag=\Pi^{\pm\pm}_\mu
   \]
   Collect these results and calculate $S[U]$ in Eq.(\ref{act})
   giving
   \[
    S[U]=S_W[U]+S_{NU}[U]
   \]
   $S_W$ is of the form of standard Wilson action for
   lattice gauge fields in LFT up to a normalization factor
   \[
    S_W[U]=-\sum_{f.p.}tr_c{\cal F}(f.p.)
   \]
   where f.p. refers to fundamental plaquette and ${\cal F}(f.p.)$ equals to one ${\cal F}^{st}_{\mu\nu}(x)$ whose
   $(-)^s\hat{\mu},(-)^t\hat{\nu}$ span this fundamental plaquette; while,
   \[
    S_{NU}[U]={1\over 2}Sp(tr_cS)={1\over 2}\sum_{x\in\intg^n}(tr_cS)(x)
   \]
   and will vanish if parallel transport $U^\pm_\mu$ satisfy
   unitarity $(U^\pm_\mu)^\dag U^\pm_\mu=\unit_c\ot\unit$.
  \section{Discussions}
  \label{sec3}
  \subsection{Non-unitary Link Variables}
   After the tedious calculation in the last section, we show that an additional term gaining contributions from those
   Wilson-loops spanning no area has to be added to the classical action of gauge fields on the lattice, if no unitarity
   of link variables is assumed. Here we illustrate the geometric significance of non-unitary parallel transports on Connes'
   distance by an one-dimensional example.
   Connes' distance on $\intg^n$ can be defined by
   $d_\Drc(x,y)=sup\{|f(x)-f(y)|:\norm{[\Drc,\pi(f)]}\leq 1\}, \forall x,y\in\intg^n$. Let $n=1,k=1$, then $\Drc, \nabla$ take on the
   forms like
   \[
    \Drc_{n=1}=\left(
     \begin{array}{cc}
      0&T^+\\T^-&0
     \end{array}
    \right),
    \nabla_{n=1}=i\left(
     \begin{array}{cc}
      0&U^+T^+\\U^-T^-&0
     \end{array}
    \right)
   \]
   In \cite{dd}\cite{ds1}, it is showed that $d_{\Drc_{n=1}}(x,y)=|x-y|$. Now consider $d_{(-i)\nabla_{n=1}}(x,y)$,
   and one can check that $\norm{[(-i)\nabla,\pi(f)]}=sup\{|U^+|^2(x)|\partial^+f|^2(x):x\in\intg\}$, noticing
   the anti-hermitian condition Eq.(\ref{her}); therefore, if $U^+$ is
   unitary, $d_{(-i)\nabla_{n=1}}=d_{\Drc_{n=1}}$, else then
   \[
    d_{(-i)\nabla_{n=1}}(x,x+k)={1\over |U^+(x)|}+{1\over |U^+(x+1)|}+...+{1\over |U^+(x+k-1)|}
   \]
   for all $x\in\intg,k=1,2,...$, namely non-unitary link variables will modify induced metric on lattices.
  \subsection{Mathematical Rigidity}
   Till now we do not apply Connes' formalism in a restrictive
   manner. In fact in \cite{rennie}, Rennie showed that what are
   recovered from Connes' axioms for commutative algebras are
   necessarily $spin^\complex$-manifolds. Hence, a lattice is
   outside to be a rigid model of Connes' formalism being
   applied to commutative algebras, unless additional structures like a {\it
   real structure}\cite{cr} or equivalently a charge conjugation in physics jargons
   is considered. As one aspect of this contradiction in our understanding, most
   constructions of \DCv operators on discrete sets including ours fail to
   fulfill {\it first order axiom} which requires $[[\Drc,f],g]=0,\forall f,g\in \Alg$. However,
   it can be shown that the ``error'' caused by this violation is
   proportional to the lattice spacing. In fact, introduce $a$ to be
   lattice spacing and still consider the previous one-dimensional
   example with $T^\pm\rightarrow T^\pm /a$ in the expression of $\Drc_{n=1}$ and
   $\partial^\pm\rightarrow (T^\pm-\unit)/a$, there is
   \eq\label{a}
    [[\Drc,\pi(f)],\pi(f^\pr)]=a
    \left(
     \begin{array}{cc}
      0&(\partial^+f)(\partial^+f^\pr) T^+\\(\partial^-f)(\partial^-f^\pr) T^-&0
     \end{array}
    \right)
   \en
   Under continuum limit, the should-be vanishing of Eq.(\ref{a}) is
   restored. As a suggestion, we would like to modify {\it first
   order axiom on lattices} to be $[[\Drc,f],g]={\cal O}(a),\forall f,g\in
   \Alg$, where ${\cal O}$ measures the convergency of operators under continuum limit.\\

   {\bf Acknowledgements}\\
    This work was supported by Climb-Up (Pan Deng) Project of
    Department of Science and Technology in China, Chinese
    National Science Foundation and Doctoral Programme Foundation
    of Institution of Higher Education in China. We are grateful
    to Alejandro Rivero for his reminding that our geometric square-root condition
    is essentially the condition for Fredholm operator with zero index.
  
 \end{document}